# A STOCHASTIC MODEL OF B CELL AFFINITY MATURATION AND A NETWORK MODEL OF IMMUNE MEMORY

T. SZABADOS[1]*, G. TUSNÁDY[2], L. VARGA[3], and T. BAKÁCS[4]

[1] Technical University of Budapest, Hungary
[2] Mathematical Institute of the Hungarian Academy of Sciences, Budapest, Hungary
[3] Hungarian Power Companies Ltd., Budapest, Hungary
[4] National Institute of Oncology, Budapest, Hungary

*(Manuscript, 1998)*

Many events in the vertebrate immune system are influenced by some element of chance. The objective of the present work is to describe affinity maturation of B lymphocytes (in which random events are perhaps the most characteristic), and to study a possible network model of immune memory. In our model stochastic processes govern all events. A major novelty of this approach is that it permits studying random variations in the immune process. Four basic components are simulated in the model: non-immune self cells, nonself cells (pathogens), B lymphocytes, and bone marrow cells (stem cells) that produce naive B lymphocytes. A point in a generalized shape space plus the size of the corresponding population represents nonself and non-immune self cells. On the other hand, each individual B cell is represented by a disc that models its recognition region in the shape space. Essential differences between self and nonself cells in the model are that (1) a negative selection filter randomly eliminates naive B cells that can bind non-immune self cells, (2) the absence of T cell help inhibits the division of such B cells. The most important random events are the attacks of a B cell against other cells whose shapes happen to be complementary to some points within the recognition region of the attacking B cell. Infection is simulated by an "injection" of nonself cells into the system. Division of pathogens may instigate an attack of naive B cells, which in turn may induce clonal proliferation and hypermutation in the attacking B cells, and which eventually may slow down and stop the exponential growth of pathogens. Affinity maturation of newly produced B cells becomes expressed as a result of selection when the number of pathogens decreases. Under favorable conditions, the expanded primary B cell clones may stimulate the expansion of secondary B cell clones carrying complementary receptors to the stimulating B cells. Like in a hall of mirrors, the image of pathogens in the primary B cell clones then will be reflected in secondary B cell clones. This "ping-pong" game may survive for a long time even in the absence of the pathogen, creating a local network memory. This memory ensures that repeated infection by the same pathogen will be eliminated more efficiently.

*Running Headline*: STOCHASTIC MODEL OF B CELL AFFINITY MATURATION

**1. Introduction.** Reviving Ehrlich's old theory of antibody formation, in 1955 Niels Jerne proposed that the information required for the production of specific antibodies preexists within the vertebrate genome. Therefore from all the possible specificities already present, an antigen selects some for the production of antibodies (see in Silverstein, 1991). Jerne's idea of "natural selection" was adapted by Burnet (1957) into the *clonal selection* hypothesis which suggested that each clone carries immunologically reactive sites corresponding in appropriate complementary fashion to one (or possibly a small number of) potential antigenic determinant. The clonal selection theory stimulated tremendously the development of experimental immunology, which seemed to confirm its most important predictions, i.e., that the immune system is able to recognize the antigens of the external environment with extreme specificity, while it avoids reaction against structures of the host. However, in the eighties more sensitive assays revealed that sera of healthy subjects and animals contain self-reactive antibodies which are produced by activated cells in normal subjects or animals (see in Bona, 1988; Stollar, 1991; Avrameas and Ternynck, 1993).

Before any experimental evidence was available, it was Jerne again who suggested in his *idiotypic network theory* (1974 a, b) that the normal immune system does react against its own components to regulate itself. Using a system of differential equations, he described a mathematical model for the normal immune system. After Jerne's groundbreaking work application of mathematics to immunology expanded tremendously. Let us summarize some important steps in this progress. (For a detailed review see Perelson and Weisbuch, 1996.) Perelson and Oster (1979) introduced the concept of general shape space to explain the paradox how a vertebrate with finitely many antibodies can recognize practically infinitely many antigens by clonal selection. For the representation of idiotopes several approaches have been used such as: bit-strings (Farmer *et al.*, 1986; de Boer and Perelson, 1991), small dimensional Euclidean parameter spaces (Segel and Perelson, 1988; Weisbuch 1990; Stewart and Varela, 1991; de Boer *et al.*, 1992), low dimensional polyhedron models (Weinand, 1990, 1991). Models were also developed to describe ontogeny (de Boer and Perelson, 1991; Stewart and Varela, 1991; Detours *et al.*, 1994), immune memory (Parisi, 1990; Weisbuch *et al.*, 1990), response to foreign antigens (Weisbuch and Atlan, 1988; Atlan, 1989; Marchuk *et al.* 1991a, b; Rose and Perelson 1992) and somatic hypermutation and affinity maturation (Weinand, 1990 and 1991; Perelson and Kepler, 1996; Pierre *et al.*, 1996). In a foreign-antigen centered immune system, (such as described by the clonal selection hypothesis) self-reactive clones must somehow be deleted to avoid autoimmunity. In a self-antigen centered, autonomous immune system (like in the idiotypic network theory) it is difficult to explain how the immune system differentiates its own structures from that of the pathogens. To solve this paradox, Coutinho (1989) postulated that a clonal, as well as a network organization coexist in the immune system. While autonomous immune network activities embody reactions against self, immune responses to external antigens are carried out by lymphocytes which are disconnected from the network.

---

* Author to whom correspondence should be addressed: Department of Mathematics, Technical University of Budapest, Egry 20-22, H ep V em, 1521, Hungary. E-mail: szabados@math.bme.hu

Mathematical modeling in immunology has been so far dominated by differential equations (most of the papers mentioned in the previous paragraph fall into this category). Another approach has been to apply automata models for simulation, see e.g. Kaufman *et al.* (1985); Seiden and Celada (1992); Celada and Seiden (1996). It would, however, be a natural idea -- like in statistical physics describing a large number of interacting particles -- to apply stochastic models in immunology too. To reason, one can recall that birth, interactions, mutations and death of the "particles" of the immune system all contain an element of chance. For instance, T and B cell receptors are assembled by a random combination of germ-line genes during the differentiation of hematopoietic stem cells. Following a random encounter with an antigen, the hypermutation process also randomly changes the receptors of immune cells. Although some models apply probabilistic approaches (see e.g. Perelson and Oster, 1979; Percus *et al.,* 1993; Merill *et al.,* 1994, Abkowitz *et al.,* 1996; Celada and Seiden, 1996), to our knowledge there has been no attempt so far to base a model of the immune system entirely on stochastic processes.

The purpose of this paper is to present a *stochastic model* which can describe some major functions of the immune system, especially B cell affinity maturation and also a possible network model of immune memory that can survive for years even in the absence of antigen. We also study random individual variations in responding to a pathogen or randomly occurring autoimmunity with a smaller population of somatic cells. Otherwise, the advantages and limitations of such a model are rather similar to that of automata models (Celada and Seiden, 1996). A stochastic model can represent several different biological components and processes simultaneously and directly and it can easily handle some biologically inherent difficulty like nonlinearity. An essential limitation is that it has to deal with a much smaller system than one would meet in reality: it is like observing a small section of a large creature under microscope. The physiological principles underlying the model are mostly based on "textbook" evidences, see Paul (1993); Roitt (1994); Alberts *et al.* (1994a, b).

**2. Principles of the Model.** A model intended to describe affinity maturation of B cells and immune memory has to address several problems.

*First*, like in natural selection, there exists neither intelligent control which would direct *genetic mutations* toward better fit, nor memory that would save cells from genetically searching a proved wrong "direction". The major effect which has physiological consequences on a B cell is the strength of antigen binding. This is like finding the source of heat in a dark room, using a single thermometer, with no direct sensing of direction and with no memory. The technique the present model applies is a stochastic search for best fit (or stochastic learning process), postulating that a B cell that can bind the given antigen stronger produces more offspring. Furthermore, the offspring may be randomly mutated, so a random variation is created in their affinity to the given antigen. Finally, when the concentration of the given antigen is decreasing, a competition arises among B cells for the antigen, and those having higher affinity would win in this *selection* process.

*Second*, an affinity maturation model has to handle the problem of *self-nonself discrimination*. Even if naive immune cell clones which can strongly bind self antigens are deleted as a result of negative selection in the thymus or in the bone marrow, still there is the danger that autoimmune clones may be produced as a result of hypermutation. In the presented model there is a three-fold defense against this danger.

(a) In accordance with physiological evidences, a *deletion filter* (representing negative selection) is applied in the model that randomly deletes *naive* B cells that can bind non-immune self antigens. The intensity of deletion decreases as the simulated immune system is aging. In the model this filter reflects an essential difference between self and nonself antigens.

(b) Although T cells are not explicitly represented in the current model, *T cell help* is. The absence of T cell help in case of B cells that react relatively strongly to non-immune self antigens inhibits their division. This is another essential difference between self and nonself in the model.

(c) Since nonself antigens which can start somatic hypermutation typically appear after birth, when the number of self cells is already very large, one can argue that at that time randomly produced self-reactive B cell clones are confronted with an overwhelming quantity of self antigens. As a result, these B cell clones would become *anergic* or *paralyzed* (Goodnow *et al.*, 1989). In the model this is simulated by the help of a nonlinear "gate function" that controls the reproduction of B cells. The reproduction process of B cells is fastest when the concentration of the complementary antigens is neither too small, nor too large. This is common for both self or nonself antigens in the model, so when nonself overgrows an upper threshold, the model immune system remains practically defenseless against it too.

As a result of the triple defense described in (a), (b), and (c), there will be "holes" in the immune system around non-immune self cells (Goodnow *et al.*, 1988 and 1990). The negative selection in the model is especially important during early ontogenesis when the smaller population of host cells is vulnerable to self-reactive virgin immune cells. As the individual reaches adult size, the large number of host cells plus the absence of T cell help can alone inhibit reproduction and affinity maturation of immune cells. Then negative selection in the model (like in reality in the thymus and in the bone marrow) becomes less essential.

*Third*, it is reasonable to expect that after a somatic hypermutation – affinity maturation process the resulted specific B cell clones may survive for a certain period of time as a *local memory*. In the model, expansion of certain B cell clones (as a result of an "infection"), under favorable conditions, stimulates the reproduction of secondary B cells which are complementary to the expanded primary B cell clones and whose receptors are, therefore, similar to the "infecting" antigens. (Of course, the similarity here means a mimicry of a binding partner and not similarity at the molecular level.) Thus a mirroring process ("ping-pong") and a local network memory develops and may last for a long time, even in the absence of the stimulating antigen. While this memory lasts, repeated infection of the same pathogen is eliminated more efficiently. This network model of immune memory essentially conforms to Jerne's immune network concept. Beside other factors, like longer living memory cells or antigen preserving follicular dendrytic cells, this could be a possible explanation of immune memory.

The physical positions of cells in the body are not reflected in the model as it is assumed that all interactions take place in a small volume of a more-or-less homogeneous liquid phase. Instead, the different non-immune self and nonself antigens (or rather, epitopes) are represented by single points in a *generalized shape space* plus by their population size, similarly as was suggested by Perelson and Oster (1979). Theoretically, the generalized shape space is a high enough dimensional space that is suitable to model all features that play an essential role in a binding between two molecules, including shape and chemical properties. Theoretical considerations compared with experimental data led to the conclusion (Perelson and Oster, 1979) that the dimension of this shape space, i.e. the number of parameters essential in describing a binding, is not too large, probably around five. (Though there are conflicting views as well as, see e.g. Carneiro and Stewart (1994)). In order to reduce the size of the model, a two-dimensional cross-section of an assumed five-dimensional shape space is considered. Furthermore, it is supposed that this cross-section is a rectangle. Finally, the problem is discretized, and the *configuration rectangle Z* in the model is a small section of the shape space, consisting of $10^6$ discrete points, being approximately $10^{-9}$th part of the whole potential immune repertoire that is considered to be about $10^{15}$ in man. Thus it is assumed that from the perspective of basic functioning, the shape space is relatively homogeneous and the possibly complicated high dimensional structure does not play an essential role.

To describe interactions and affinity maturation of B cells, they are represented by a *recognition region* (in a two-dimensional case it is in fact a disc), and not by a single point in the shape space. The recognition region can be defined by the coordinates of the center and by the radius of the disc. A B cell can recognize and bind any antigen that is characterized by a point in the shape space complementary to some point in the B cell's recognition region. During hypermutation and affinity maturation the recognition discs of some B cells may get closer to a complementary point (i.e., complementary shape) of the antigen, which means larger affinity (binding strength) between antigen and B cell receptor. Hence during a mutation-selection process a focusing chain reaction may develop, concentrating on the given antigen. We also investigated a model where the radius of the recognition disc (i.e., the magnitude of the antigen repertoire the B cell can react) also randomly changed as a result of somatic hypermutation. Since we do not know of experimental evidences of such phenomenon, these results are not shown in this paper. Here we just mention that then the process of focusing on the given antigen is more effective, more accurate, as the B cells with smaller and fitter repertoire overcome other B cell clones during selection.

The *stochastic* nature of the model means that stochastic events govern cell birth, reproduction, mutual encounter and death. Each stochastic process applied in the model is a Markov process with a discrete state space, mostly a birth-and-death process. In such a case the waiting times between two events are exponentially distributed and are conditionally independent. At each moment during a Monte Carlo simulation there is a set of competing independent random events whose waiting times have exponential distributions. (Therefore each applied distribution is characterized by a single parameter, the expectation.) Only the first of these random events will be realized, and all influenced waiting times will be re-generated, usually with new expected values incorporating the effect of the actual event. In general, the expectation of any waiting time may depend on the actual state of the whole system. (These expected times will be denoted by $\tau$ plus a proper subscript in the sequel.) A simulation typically starts a few days after "conception" and lasts until several hundred days after "birth", the typical unit of time corresponds to about one tenth of a day.

There are four basic components in the present model:

(i) *Non-immune self cells* (self cells or self antigens, for short). Each specific type is represented by a single point in the shape space and by its population size. Some cells of the population are born at the start of the simulation (typically at a time instant representing one of the first days of pregnancy), while others are born as offspring at random time instants later. During their lifetime immune cells continuously attack them, though several factors, mentioned above, may restrict these attacks.

(ii) *Nonself cells* (nonself antigens). Again, each specific type is represented by a single point in the shape space and by the size of its population. The time when a specific type (e.g. a pathogen) enters the system is arbitrary, but typically it happens at time instants representing "after birth" state, when the number of self cells is already very large and the immune system is ready to fight. They also reproduce at random time instants and are continuously attacked by immune cells.

(iii) *Bone marrow cells* (stem cells). The sole role of the bone marrow in the model is to function as a biological clock: to measure the size of the body. This means that the number of bone marrow cells controls the birth rate of naive immune cells, the maximal "permitted" number of lymphocytes, and the intensity of the negative selection filter. The bone marrow is represented by the size of the population. Again, some cells of the population are born with a small time delay after the start of the simulation, while others are born in a random reproduction process later.

(iv) *B cells* (immune cells). Each B cell is represented individually by its recognition disc in the model. *Naive* immune cells are continuously produced in the bone marrow at random time instants. Their position in the shape space is randomly chosen by uniform distribution, correspondingly their affinity to a specific antigen is typically small. Each B cell, independently of the others, at random time instants attacks self, nonself, or immune cells whose positions in the shape space are complementary to some point in the B cell's recognition disc. A possible result of a successful attack is the birth of new (*committed*) B cells. For simplicity, only B cells are represented explicitly in the model, assuming that the number of antibodies in a certain clone and the number of corresponding T cells are proportional to the number of B cells of this type. A more accurate model describing antibody formation was given e.g. by de Boer and Perelson (1991).

**3. Births of Cells.** In the model it is typical that when the size of certain cell population gets larger the per capita birth rate of this cell type decreases. Thus the size of the clone first increases exponentially, later it slows down, and in many cases gets relatively stable, depending on model parameters. To control the birth rate we use a class of "gate" functions, previously applied by many other authors (see e.g. Kaufman *et al.*, 1985; de Boer and Perelson, 1991):

$$g_{\theta,\eta}(x) = \frac{\theta^\eta}{\theta^\eta + x^\eta} = \left(1 + \left(\frac{x}{\theta}\right)^\eta\right)^{-1} \quad (x \geq 0; \theta > 0, \eta > 0), \tag{1}$$

where $x$ denotes the independent variable (e.g. the number of cells) and $\theta$, $\eta$ are parameters. Formula (1) describes a decreasing function, which is equal to 1 for $x = 0$, 1/2 for the threshold value $x = \theta$, and decreases to 0 as $x \to \infty$. In one case in the sequel we need a composite "gate" function

$$\tilde{g}_{\theta_1,\theta_2,\eta}(x) = \left(1 - g_{\theta_1,\eta}(x)\right) g_{\theta_2,\eta}(x) = \frac{x^\eta}{\theta_1^\eta + x^\eta} \frac{\theta_2^\eta}{\theta_2^\eta + x^\eta}, \tag{2}$$

where $x \geq 0$, $0 < \theta_1 < \theta_2$, and $\eta > 0$. Formula (2) describes a bell-shaped function in logarithmic $x$ coordinate. When $\theta_1 \ll \theta_2$, the value of $g$ for $x = \theta_1$ or $x = \theta_2$ is about 1/2, while at the maximum point $\sqrt{\theta_1 \theta_2}$ it is close to 1. Both in (1) and (2), the larger the value of the parameter $\eta$, the more square like these functions become.

(i) *Births of new somatic self cells.* In this case,

$$\tau_s = \frac{\tau_{s0}}{s\, g_{\theta_s,\eta_s}(s)} = \frac{\tau_{s0}}{s}\left(1 + \left(\frac{s}{\theta_s}\right)^{\eta_s}\right), \tag{3}$$

where $\tau_s$ is the conditional expected waiting time between two births, $\tau_{s0}$ is its initial per capita value. Further, $s = s(z,t) \geq 1$ is the actual number of the given type somatic self cells, where $z$ is a point of the configuration rectangle $Z$ that actually carries a number of self cells at time $t \geq 0$. Typical values of the parameters are $\tau_{s0} = 300 - 500$, $\theta_s = 1400$ and $\eta_s = 4$. (Time in the model is measured in one tenth of a day.) Formula (3) indicates that when the number of cells $s$ becomes significantly larger than the threshold value $\theta_s$, the per capita birth rate $\lambda_s = 1/(s\tau_s)$ gets close to zero.

(ii) *Births of new nonself cells.* An arbitrary quantity of nonself cells of a given type can enter the model at an arbitrary time instant. This effect is called infection. When the number of cells of the given type becomes larger, the per capita birth rate decreases. Similarly as with self cells, we assume that the conditional expected waiting time between two births is given by

$$\tau_n = \frac{\tau_{n0}}{n\, g_{\theta_n,\eta_n}(n)} = \frac{\tau_{n0}}{n}\left(1 + \left(\frac{n}{\theta_n}\right)^{\eta_n}\right), \tag{4}$$

where $\tau_{n0}$ is its initial per capita value and $n = n(z,t) \geq 1$ is the actual number of cells in the given nonself population. Here $z$ is a point of the configuration rectangle $Z$ that actually holds a number of nonself cells at time $t \geq 0$. Typical values of the parameters are $\tau_{n0} = 30 - 250$, $\theta_n = 7000$ and $\eta_n = 2$.

(iii) *Births of new bone marrow stem cells and new naive B cells in the bone marrow.* This process has two components: the reproduction of bone marrow cells ($\tau_m$) and the production of virgin immune cells in the bone marrow ($\tau_i$). In both cases, the expected waiting times between two births depend on the actual number $M = M(t) \geq 1$ of bone marrow cells. Thus the per capita birth rates decreases as the number of bone marrow cells increases:

$$\tau_M = \frac{\tau_{M0}}{M\, g_{\theta_M,\eta_M}(M)} = \frac{\tau_{M0}}{M}\left(1 + \left(\frac{M}{\theta_M}\right)^{\eta_M}\right) \tag{5}$$

and

$$\tau_i = \frac{\tau_{i0}}{M\, g_{\theta_i,\eta_i}(M)} = \frac{\tau_{i0}}{M}\left(1 + \left(\frac{M}{\theta_i}\right)^{\eta_i}\right), \tag{6}$$

where $\tau_{M0}$ and $\tau_{i0}$ are the initial per capita values. Typical values of the parameters are $\tau_{M0} = 400$, $\theta_M = 140$ and $\eta_M = 4$, respectively, $\tau_{i0} = 250$, $\theta_i = 300$ and $\eta_i = 1$. Substituting the parameter value $\eta_i = 1$ into (6), it follows that as the value of the number of bone marrow cells $M$ approximately stabilizes in time, the birth rate of naive immune cells also becomes nearly stable, since then $\tau_i = \tau_{i0}\left(M^{-1} + \theta_i^{-1}\right)$.

**4. Encounters of B Cells.** The most important random events that determine the dynamics of the model are the encounters and hypermutation of B cells. To explain these, let us have a closer look at the configuration rectangle $Z$ whose geometry is intended to model some important characteristics in immunology. The $x$ and $y$ coordinates of any point are integer numbers between $-x_{\max}/2$ and $x_{\max}/2$, where $x_{\max}$ is a specified constant (e.g. $10^3$). An epitope or (the center of) a B cell receptor is represented by a point $z = (x, y)$ in the configuration rectangle and is considered to be *exactly complementary* to an epitope or B cell receptor represented by the point $\bar{z} = (-x, -y)$, symmetric to the origin. This convention was used by de Boer *et al.* (1992) as well. Instead of the Euclidean distance we use maximum distance in the model:

$$\|z_1 - z_2\| = \max\left(|x_1 - x_2|, |y_1 - y_2|\right) \tag{7}$$

where $z_k = (x_k, y_k)$, $k = 1, 2$. The "circles" in this metric are squares with sides parallel to the coordinate axes. Further, we use torus topology, i.e., we identify the upper and lower edges, also, the left and right edges of the rectangle, in order to eliminate boundary effects. Our simulations *without* torus topology showed that except for the neighborhood of boundary points, the major behavior of the model remains the same.

A B cell is determined by the following parameters: its *center* $z_b = (x_b, y_b) \in Z$ and its integer valued *recognition radius* $r_b$, where $r_b = r_0$ is a given constant (e.g. 160). These parameters define the recognition region (a square) in whose reflected image (to the *x*-axis) the given B cell is capable to bind an antigen. The set of all points $(x, y)$ in the configuration rectangle whose distance from $(-x_b, -y_b)$ is less than or equal to $r_b$ (that is, the set of antigens recognizable by the given B cell) will be called the *object region* of the B cell.

Newly born *naive B cells* are randomly, uniformly spread over the configuration rectangle. The length of the lifetime of a B cell is also random, with the same expectation $\tau_l \approx 140$ (about 14 days), either it is naive or committed. This is longer than the generally accepted life-span of a B cell (say 2 days), but it is explained by the circumstance that antibodies are not explicitly represented in the model (their average life-span is around 20 days). The birth rate, the life-span, and the radius of the recognition ball of naive B cells are chosen so that the object regions of the naive population can cover most of the configuration rectangle with large probability at any time after birth.

The *affinity* of a B cell to a given antigen is represented by the distance *d* between the target and the center of the object region of the B cell, and which will be called now on the *distance of attack*. Concretely, the affinity $\alpha(d)$ is calculated in the model as

$$\alpha(d) = g_{\theta_d, \eta_d}(d) = \left(1 + \left(\frac{d}{\theta_d}\right)^{\eta_d}\right) \quad (0 \leq d < r_b), \tag{8}$$

where typically, $\theta_d = 100$ and $\eta_d = 1$. Substituting these parameters into (8), the formula simplifies as

$$\alpha(d) = (1 + 0.01d)^{-1} \quad (0 \leq d < r_b). \tag{9}$$

Of course, the affinity is larger if the fit is better, i.e., when the distance of attack *d* is smaller. The range of affinity is much more compressed in the model than in reality, as the whole model itself (cf. Section 6 below).

Naive B cells are subject to a random deletion filter (*negative selection*) which can destroy immune cells that are capable to strongly bind to some self antigen. Each freshly born naive immune cell is "tested" in the model: if its recognition region contains some self antigens and its affinity $\alpha(d_{\min})$ to the closest self antigen is larger than a given constant $\alpha_0$ (e.g. 0.15), the new immune cell gets into the list of "dangerous immune cells". At random time instants one such cell is randomly chosen and destroyed. The expectation of waiting time between two actions of the random filter depends on the number of bone marrow cells, its per capita intensity again decreases with "aging", i.e., when the number of bone marrow cells $M$ gets larger:

$$\tau_f = \frac{\tau_{f0}}{M\, g_{\theta_f, \eta_f}(M)} = \frac{\tau_{f0}}{M}\left(1 + \left(\frac{M}{\theta_f}\right)^{\eta_f}\right), \tag{10}$$

where $\tau_{f0}$ is the initial per capita expected waiting time. Typical values of the parameters are $\tau_{f0} = 200$, $\theta_f = 300$ and $\eta_f = 1$.

Each B cell becomes active after an exponential waiting time, independently of other B cells. A typical value of the expected waiting time between two actions of a single B cell in the model is $\tau_b \approx 5$ (i.e. about 0.5 days). Each time when a B cell is active, it randomly selects an antigen currently in its object region (if there is any) and attacks it. The target antigen can be a self or a nonself cell (which are represented by one point in the recognition rectangle), or a B cell (which is represented by its center in this case). The probability of selecting a certain antigen is proportional to the affinity $\alpha(d)$ between the antigen and the given B cell, where $d$ is the distance of attack. The larger the affinity between the antigen and the B cell, the larger the antigen's chance to be selected as in this case it has larger chance to attach to and stimulate the B cell. If the randomly selected target is a self or nonself cell, one individual from that population is killed. B cells when selected are destroyed only with a certain probability $p_d \approx 0.01$. The reason behind this is that in the current model a B cell represents a number of corresponding antibodies too, and when a B cell attacks another B cell, in fact it could be only an antibody of that type. Then that antibody is destroyed, but not its parental B cell.

**5. Proliferation, hypermutation, and affinity maturation.** After a successful attack of a B cell, a proliferation and hypermutation process may begin in the model. The recognition regions of the offspring of an activated B cell can be different from that of the parental cell and from each other. This represents a random somatic hypermutation. The probability of mutation per cell per reproduction in the model was taken to be between 0.01- 0.2. This mutation rate is significantly larger than the experimentally observable somatic hypermutation rate (about $10^{-3} - 10^{-4}$ per cell per reproduction), but because of the relatively small, compressed number of B cells in the model (cf. Section 6 below), mutations otherwise would be practically absent.

*Affinity maturation* and the corresponding focusing on a certain antigen are made possible by somatic hypermutation and a consecutive selection process in the model. Suppose that the given parental B cell has center $z_b \in Z$ and radius $r_b$ (typically 160 in the model) and some mutation occurs at a division. On the one hand, the square over which the offspring of this B cell are (randomly, with uniform distribution) spread is centered at $z_b$, so the basis from which mutation starts is the current genetic state of the parental B cell. This means that for the center $z_{new}$ of an offspring B cell we have the inequality $\|z_{new} - z_b\| \le r_{sp}$, where $r_{sp} = 20-80$. However, since the center of offspring is selected at random in the square given above, it could be as likely closer to the target than the parent is, as farther from it. Either way, there is a chance that an offspring will fit better than its parent does, and then during a selection process later it can overcome other, less fit competitors.

The most important and most complex part of the dynamics is the formula that determines the *number of offspring* (or the number of divisions) resulting from a given encounter of a B cell. This formula essentially corresponds to a growth differential equation commonly used in the literature (Jerne, 1974b; Perelson, 1989; Weisbuch, 1989), to mention just a few examples. The number of offspring $b_{new}$ is determined by five factors:

$$b_{\text{new}} = k_b \, \alpha(d) \, \tilde{g}_{\theta_{c1},\theta_{c2},\eta_c}(c) \, g_{\theta_B,\eta_B}(B_T)\left(1 - g_{\theta_d,\eta_d}(d_{\min})\right) \qquad (11)$$

(rounded to the nearest integer). Here $k_b$ is a constant (typically around 2-3) that gives the maximum number of offspring that may result from a given encounter. The functions $g$ and $\tilde{g}$ are defined by (1) and (2), respectively, and $\alpha(d)$ denotes the affinity between the B cell and the given antigen, defined by (8). Among the variables, $d$ is the distance of attack and $c$ denotes the actual local concentration (number) of cells in the object region of the B cell. Further, $B_T$ is the actual total number of B cells in a neighborhood of the given B cell and $d_{\min}$ is the distance between the center of the object region of the B cell and the closest self antigen. The number of offspring $b_{\text{new}}$ could be zero too, or even larger than 1. In the latter case the produced offspring are born one after the other with a random time delay with expectation $\tau_b / b_{\text{new}}$. The random length of lifetime of each new B cell has the same expectation $\tau_l$ as in case of a virgin B cell. (This means that e.g. longer living memory cells are not represented in the current model.)

Let us see the significance of each factor of (11) in detail. The first factor, the maximum number of offspring $k_b$ is chosen to be larger than 1, because it is multiplied by four factors ranging between 0 and 1, and even so typical values of the actual number of offspring $b_{\text{new}}$ are 0 or 1. If $b_{\text{new}}$ happens to be, say, 2, then it represents two divisions of the B cell combined.

The second factor, the affinity $\alpha(d)$ has the effect that when the affinity (i.e., the strength of attachment) between target and B cell receptor is large, an induced division of B cell is more likely.

The third factor depending on the local concentration (number) $c$ of antigens in the B cell's object region is a sort of switch. If the concentration is small (compared to the lower threshold $\theta_{c1}$, which can be 30 e.g., and on the value of $\eta_c$, which can be 2, say), this switch is turned off. This is meant to inhibit the effects of random fluctuations to the reproduction process of B cells. Of course, there is an inverse relationship between the stability and the sensitivity of such a dynamical system (Perelson, 1989). If the value of the lower threshold $\theta_{c1}$ is larger, it may inhibit the immune process, the immune system becomes less sensitive to nonself antigens, and it may block the emergence of immune memory as well. While if this lower threshold is too small, then a percolation process of B cells can start without any external stimulus. On the other hand, if the antigen concentration is very large (compared to the upper threshold $\theta_{c2}$, which can be 500 e.g., and also on $\eta_c$), the switch is turned off as well: then the B cell becomes anergic (Goodman *et al.*, 1989; de Boer and Perelson, 1991). This is an important factor against autoimmunity, because the large concentration of self cells in itself is a defense against the activation of immune cells. The number of offspring is largest when the value of antigen concentration is between the lower and upper thresholds, near to the maximum point $\sqrt{\theta_{c1}\theta_{c2}}$ of the function $\tilde{g}$.

Further, the fourth factor depends on the total number $B_T$ of B cells in a neighborhood of the B cell with a given radius $r_T$ (= 240, say). The effect of this factor is (as explained by energetic reasons and the homeostasis of immune cells) that the number of immune cells in a given neighborhood cannot increase "without bound", and an increase should be the consequence of a strong enough exterior stimulus. In other words, this is a non-specific inhibition concerning reproduction of B cells; similar to the one used by Perelson (1989) and Merill *et al.* (1994). Typical values of the parameters are $\theta_B = 0.3M$ and $\eta_B = 2$, where $M$ is the actual number of bone marrow cells. In other words, the size of the bone marrow population (which measures the size of the body in the model) sets a limit on the total number of B cells in a neighborhood.

Finally, the fifth factor represents T cell help (more accurately, the absence of T cell help). This factor is small, when the "shape-distance" $d_{\min}$ between the center of the object region of the B cell and the nearest self antigen is small compared to the threshold value $\theta_d \approx 100$. In other words, if the shape of the typical antigen this B cell clone can process resembles some self antigen too much, there is a smaller chance for division of this B cell, since then T cell help does not exist or small.

**6. Simulation Results.** Suppose that the generalized shape space is a hypercube in a 5-dimensional space. Only a discrete grid is considered with non-negative integer coordinates in the hypercube, say, $0 \leq x_i \leq 10^3$ ($i=1,...,5$). Then altogether there are $(10^3)^5=10^{15}$ discrete points in the shape space, that is considered as the potential immune repertoire. Fixing a two-dimensional coordinate subspace, there are $(10^3)^3=10^9$ disjoint cross-sections with integer coordinates, parallel to the fixed subspace. The configuration rectangle is one such randomly chosen two-dimensional cross-section with $10^6$ points.

Assuming $10^{13}$ self cells in a newborn at birth, with about $10^5$ different types of antigens, one has roughly $10^8$ of a typical one. Out of the $10^5$ different types, the configuration rectangle as a $10^{-9}$ portion of the whole space, will typically hold at most a few specific types of self cells. To alleviate the handling of dynamics in the model another dimensional compression is applied too. Imagine that, say, $10^8$ individual self cells of a given type are spread on a small 5-dimensional hypercube grid. So $(10^8)^{2/5}$, i.e. of the order $10^3$, self cells of this type are in the two-dimensional cross-section around birth. One similarly obtains that a few days after conception, when the number of self cells of the given type is around $10^3$ only, the configuration rectangle contains of the order 10 cells of this type. Hence in our simulations the number of self cells of a specific type increases from around 10 to around $10^3$ with typical reproducing cycle 30-60 days. This way one can compress the dynamics of reproduction.

In the case of nonself cells the situation is similar: the configuration rectangle contains at most a few specific types and $10-10^3$ individual cells from each. Because of the absence of explicit antibodies, the simulated immune system can fight only slowly dividing nonself cells, having reproducing cycle 3-10 days typically.

Turning to the case of B cells, the typical recognition volume that a (naive) lymphocyte can cover in the shape space is about a $10^{-5}$ fraction of the whole volume (Segel, 1985). In the present 5-dimensional case it means a radius one tenth of the total size 1000, i.e. about 100. Supposing that a newborn has $10^{11}$ B cells, the selected cross section (about $10^{-9}$ portion of the whole shape space) will typically hold about 100 of them. In our simulations the average lifetime of a B lymphocyte is around 14 days. A stimulated B cell is able to divide once per day as an average.

The time unit corresponds to approximately one tenth of a day. The simulation typically extends from the first days of pregnancy to about 100-700 days after birth. The configuration rectangle contains only a few (typically one or two) specific types of self cells. At the beginning their cell population is small (say 20). Since during early embryonic life cell cycles are between 8 to 60 min. (similar to that of the pathogens, Alberts *et al.*, 1994a), the host cells may reach a critical number before the immune system emerges. In our model we simulated this kinetics by varying the cell division time, i.e. cell cycle was typically twenty times faster at the beginning than at the end of simulation (see formula (3) above), so at the beginning the growth is practically exponential.

First let us consider a typical simulation with one type of self antigens and a single infection after birth (Figs. 1-4). The immune system starts to build up a few days after the start of the simulation. The negative selection of those immune cells that can strongly bind self cells starts at the same time. At birth the immune system is relatively small, mostly consists of naive polyreactive, immune cells (about 50). The number of self cells is of the order $10^3$ then. As a result of negative selection, the absence of T cell help, and the large concentration of self cells, the concentration of the "dangerous" immune cells that could bind self cells is less than the concentration of immune cells in other "non-dangerous" regions of the configuration rectangle. (So there will be "holes" around the complementary shapes of self cells.) After birth, the birth rate of self cells is much slower, also the activity of negative selection is significantly smaller. (The absence of T cell help and the large concentration of self cells are satisfactory defense alone.)

Several days after birth a quantity of nonself cells (say 100) is placed on an arbitrary point in the configuration rectangle, representing an infection. First the non-specific immune cells that are able to interact with the given type of nonself cells "realize" the increase in the concentration in their object region compared with their neighborhood. They begin to inhibit the exponential growth of the number of nonself cells. As a result, the increase of the foreign population slows down (Fig. 1), or remains almost constant (not shown). Without immune defense infection would grow exponentially (Fig. 5). At the same time a clonal selection process of B cells starts together with affinity maturation based on hypermutation. We assume in our model that hypermutation and affinity maturation starts already during the primary immune response (Kocks and Rajewsky, 1989), although first it is slow. This process becomes more effective after several days, when the number of nonself cells begins to decrease and the nearly complementary B cells start "fighting for the prey". In the meantime, the total B cell population becomes 3-5 times larger than originally (Fig. 2), mostly due to the increase of antigen specific B cells (Fig. 3). Then, as a result of the fast decrease in the nonself population, B cells in the neighborhood of antigen "realize" the large concentration of antigen specific B cells. They start to divide and focus on the antigen specific immune cells. Thus (under favorable conditions) a "mirroring" process is created and there will be two symmetrical subregions of the configuration rectangle with highly elevated concentration of B cells: one population has complementary shape to the original antigen and another population has similar shape as the original antigen (Fig. 4). These two sub-populations may keep stimulating each other ("ping-pong"). This local network memory may last from several days to a hundred years in simulated model time, even in the absence of the pathogen. While memory persists, a repeated infection of the same type is eliminated faster and more efficiently than the first one (Fig. 6).

Of course, a stochastic process shows random fluctuations. Repeating the experiment shown on Fig. 1 a hundred times with the same parameters, but with different random numbers, we had six occasions when the pathogen population outgrew the host. In the remaining 94 cases the mean eradication time of the pathogen was 4.76 days with 2.04 days standard deviation in simulated model time.

The capacity of this model immune system, similarly to real life, is limited. Large dose of and/or fast dividing nonself cells will outgrow the host like in Fig. 5. However, the quantity of the nonself cells and their division rate play an important role in the formation of memory. A small quantity of antigens or slowly dividing cells are eliminated very fast, and as a result, the population of the immune cell clones with complementary shape to the antigen will be smaller. More importantly, the population with similar shape to the original antigen will be even smaller and the resulting memory could be very weak and short living. Several inner parameters of the model influence the formation of memory as well. Probably the most important is the B cell capacity given by $\theta_B$ in (11). While with the parameters given above the model gave long lasting memory in about 80% of the experiments, decreasing B cell capacity by 30%, long lasting memory was formed only about 50% of the time.

The model immune system was tested for uniformity: an "infection" of the same size and division rate was placed on 100 points of a grid over the entire configuration space, one-by-one, starting from the same initial conditions and same random numbers (Fig. 8). The eradication time, as expected, was not uniform: the pathogen proliferated "with no bound" in the close neighborhood of the host cells (due to the "holes" of the immune system there).

It is also important to note that, by accident with small probability, autoreactive clones may emerge during embryonic life and if they happen to become activated, then the host can be killed (Fig. 7).

It has been suggested that B cells might play a role in autoimmune diseases either directly via the production of autoantibodies or indirectly by acting as efficient antigen-specific accessory cells leading to T cell autoreactivity (Cash *et al.*, 1996). Autoimmunity may occur by two, not mutually exclusive, mechanisms in a "local" environment of our model. Suppose that a virus masks self antigens by presenting its peptides, thus most autoreactive immune cells are freed from continuous self-antigen binding and a focusing reaction may start against the remaining uninfected cells. The reason for this is, as we assume, that the number of uninfected self cells falls below the critical activation threshold level, provoking an immune attack (see the fourth factor depending on the concentration difference in (11)). While the infection is being brought under control, simultaneously an autoimmune attack develops against the *uninfected* host cells (Fig. 9). Similar autoimmune reaction may also start when a virus infection modifies the sterical - chemical properties (in the model: the position in the shape space) of a minority of self cells.

Although our model is meant to describe the physiological behavior of B cells, we attempted to simulate some basic observations in transplantation immunology too. It is well documented that immune attacks in a graft-versus-host (GVH) reaction after allogeneic bone marrow transplantation are directed principally against a minority cell population. These are the fast renewing areas in the blood, epithelial lining of small intestine, liver or in the basal layer of the skin (Ferrara and Deeg, 1991; Ferrara *et al.*, 1996). Since these areas contain a high proportion of undifferentiated stem cells, it has been suggested that these are the main targets of lymphocytes. This is consistent with the prediction of our model: the large number of resting allogeneic cells is a sufficient defense against immune attack, while the differentiating cells are in minority, thus they are not "immune" against attacks. The transplantation reaction against an organ allograft may be a similar case in point. The number of invading cells often seems insufficient to mediate rejection by cell-to-cell contact with every allogeneic target. This finding has supported the claim that certain critical elements of the graft, such as its blood vessels, are the actual site of graft destruction (Paul, 1993; Mayer *et al.*, 1988).

To simulate such a scenario, an immune reaction was induced by "switching off" the negative selection of autoreactive clones at time 3000. (It is assumed that after bone marrow transplantation the self cells are not defended by negative selection.) The fate of two populations of host cells, occupying different places in the configuration space and having different maximal sizes was followed (Fig. 10). Since the smaller population could not reach a critical number by the time when the negative selection was switched off, they activated the autoreactive clones and were attacked. Consequently they stopped dividing and by time 3500 their number decreased. The cells in the larger population continued to grow, despite the fact that the negative selection was switched off since its size was already above the critical upper threshold for lymphocyte activation to occur.

**7. Conclusions.** The presented stochastic model tries to address some major questions of immunology, especially problems that concern B lymphocytes: self-nonself discrimination, response to nonself antigens, B cell hypermutation with affinity maturation, establishment of local memory and more efficient secondary response to an antigen, prevention of autoimmunity. To our knowledge this is the first model in which all events are governed by *stochastic processes*. There is a special emphasis on the investigation of random variations which is a main novelty of this model. Summarizing, the following characterize the present model: (i) naive immune cells cover most of the configuration space, (ii) immune cells progress autonomously under the influence of their environment, (iii) under suitable conditions B cells stimulate the reproduction of complementary or near complementary B cells (specific stimulation), (iv) the total number of B cells has a limiting effect on B cell division (non-specific inhibition). It may be stated that the model represents an *immune network*, similar to the one described by Jerne (1974a). It can be stated too that with any realistic immune system there is a chance for random errors and there is large room for random individual variations.

The presented model concentrates on the above-mentioned main questions, and works with a very limited number of specific elements: self and nonself antigens, B cells and bone marrow cells. It is interesting that such a relatively simple stochastic model can describe the above-mentioned features of the immune system qualitatively quite well. The absence of several important elements of the immune system, especially non-specific immune cells, antibodies and explicit T cells, may explain that the model cannot fight nonself cells that divide realistically fast. Of course, simplifications about the generalized shape space, antigen-receptor binding, etc. may contribute to inaccuracies too. However, we hope that the simulation results presented above can convince the reader that entirely stochastic models can reflect some major phenomena in immunology, and this way they can supplement other, more widely used approaches, like differential equations or automata models.

**Acknowledgment.** We thank Alan S. Perelson for sending us some of his (then) unpublished manuscripts, M. Farkas, G. Michaletzky, Lídia Rejtő, and David Segal for useful comments and suggestions. Our thanks are due to Attila Czégeni, Péter Fekete, and Gábor Szabados for their help in preparing the manuscript. This work was supported in part by Hungarian National Scientific Research Foundation (OTKA) grants no. T 16237, T 6362, and T 15668.

**Figure legend**

Figure 1: Simulation of elimination of pathogens by the model immune system. (The unit of time is about one tenth of a day.)

Figure 2: Random distribution of B cells in the configuration rectangle just before "injection" of nonself cells at $t = 3000$, from the simulation shown in Fig. 1. Position of self cells: x, nonself cells: ∗. (a) B cell centers (b) B cell recognition regions.

Figure 3: Clonal expansion of B cells, complementary to the pathogens, in the configuration rectangle at $t = 4000$, from the simulation shown in Fig. 1. (a) B cell centers (b) B cell recognition regions.

Figure 4: "Mirroring": clonal expansion of B cells, complementary, respectively, similar to the pathogens, in the configuration rectangle at $t = 5000$, from the simulation shown in Fig. 1. (a) B cell centers (b) B cell recognition regions.

Figure 5: Simulation of pathogen growth in the absence of immune defense.

Figure 6: Immunization: simulation of elimination of repeated infections.

Figure 7: Simulation of a small probability activation of an autoreactive B cell clone, which kills the fetus.

Figure 8: Uniformity analysis of eradication time of 20 pathogens over the entire configuration space.

Figure 9: Simulation of an autoimmune reaction by switching off the negative selection of autoreactive clones at $t = 3000$: a small population is deleted, while a large population survives.

Figure 10: Simulation of an autoimmune reaction against an uninfected cell minority, following a viral infection.

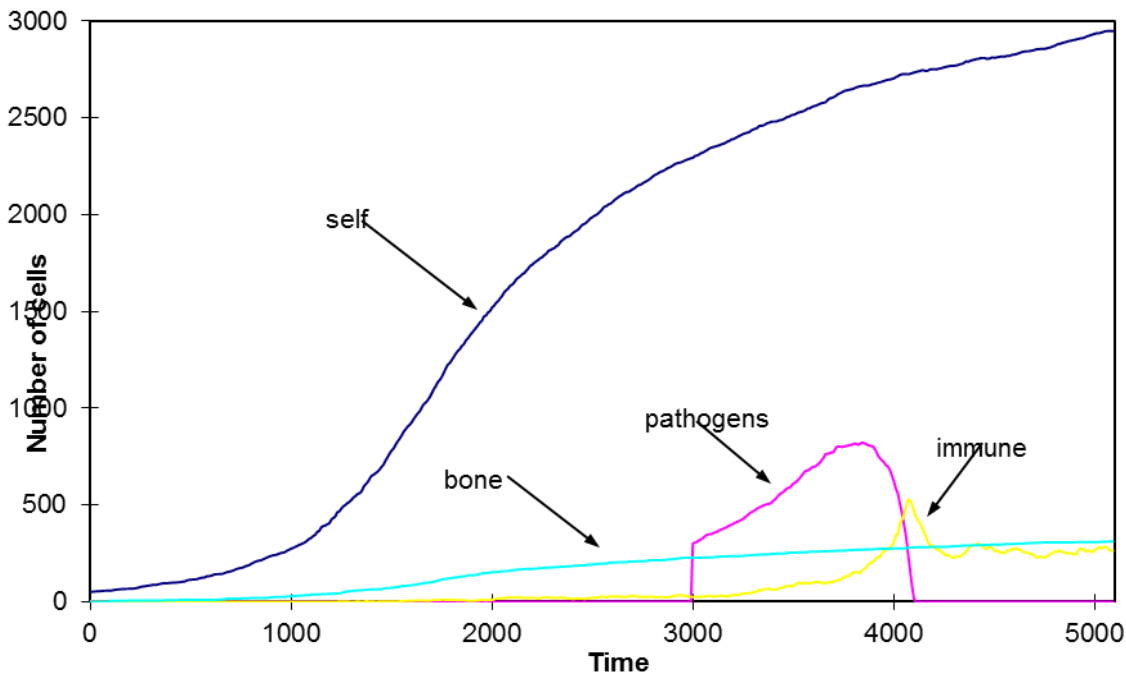

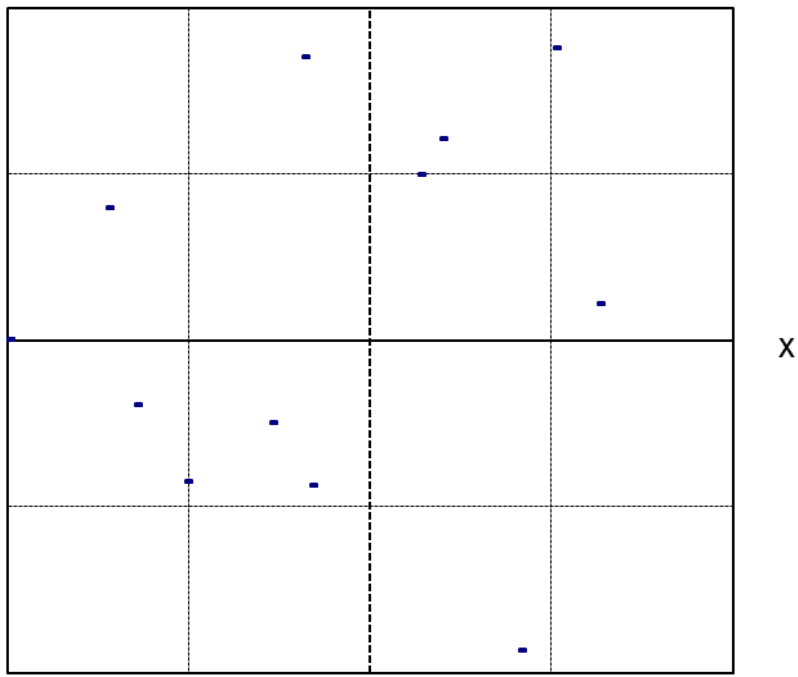

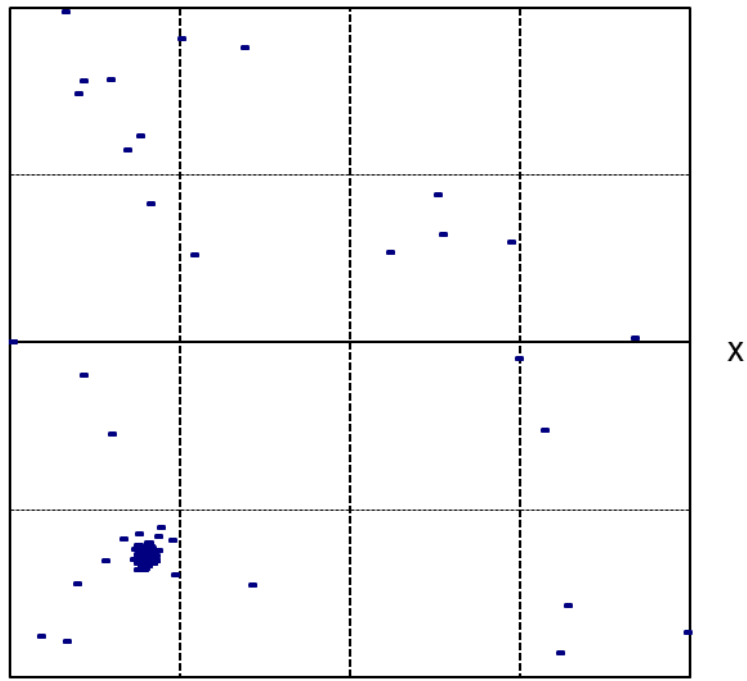

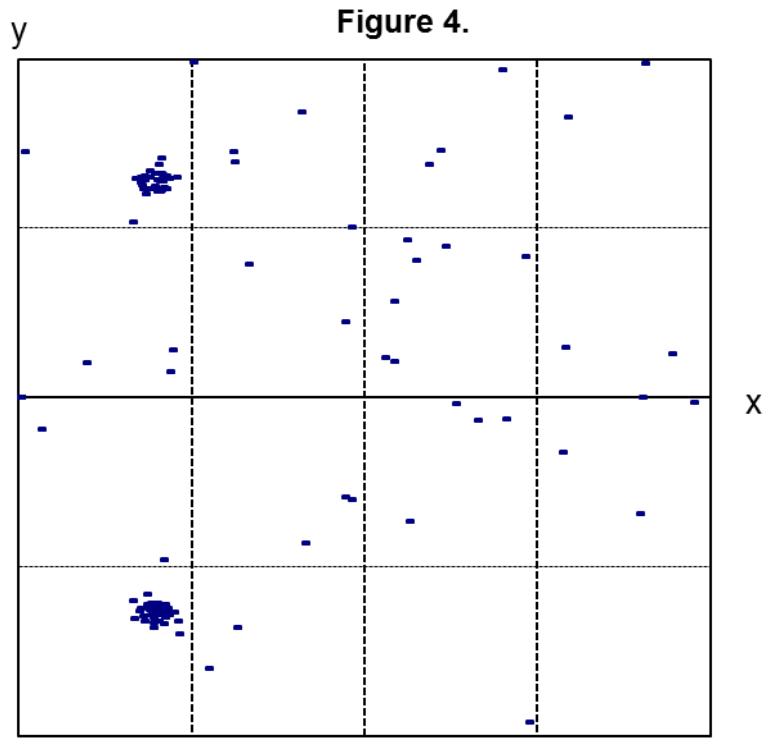

**Figure 4.**

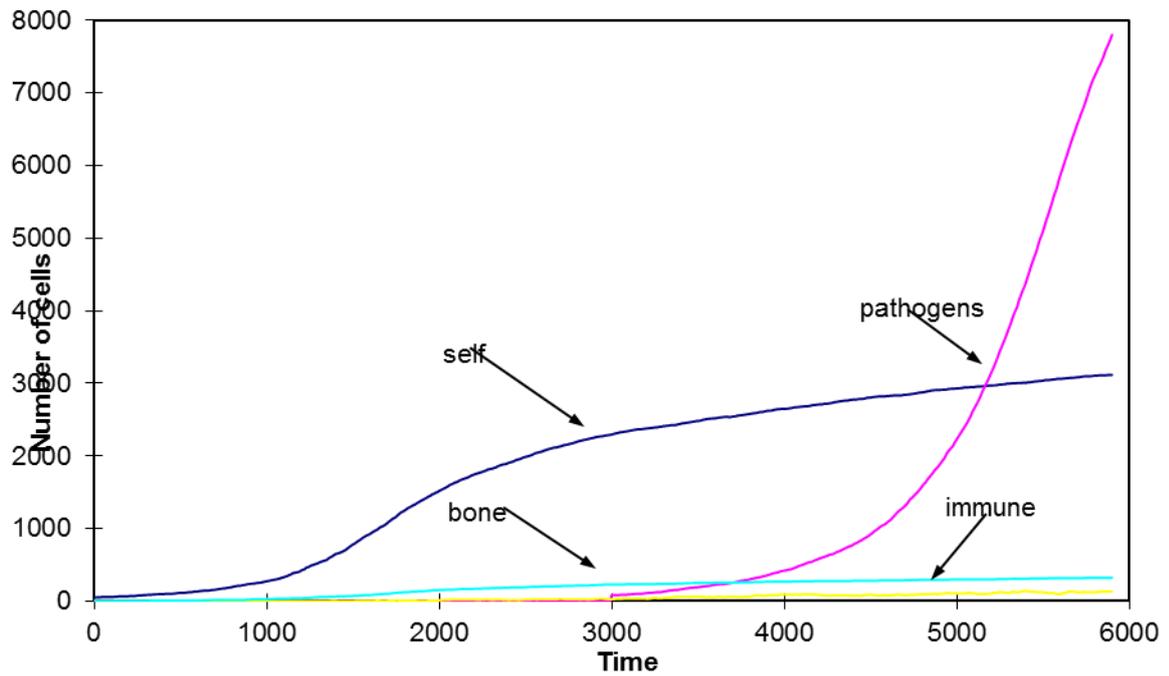

**Figure 5.**

**Figure 6.**

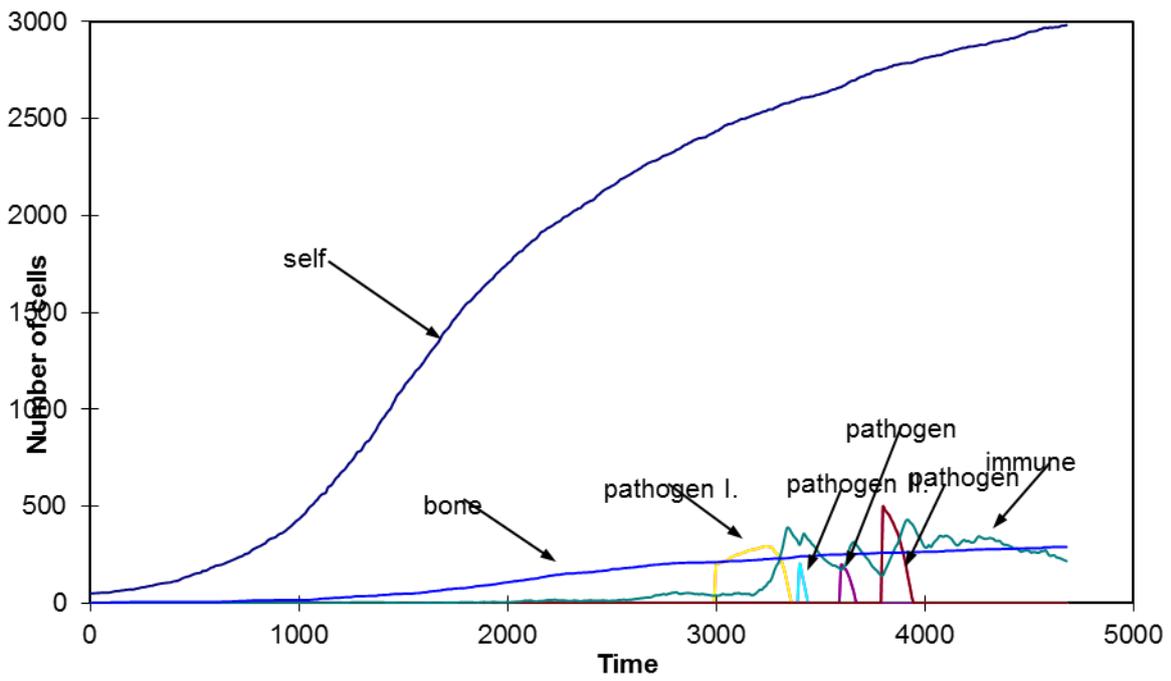

**Figure 7.**

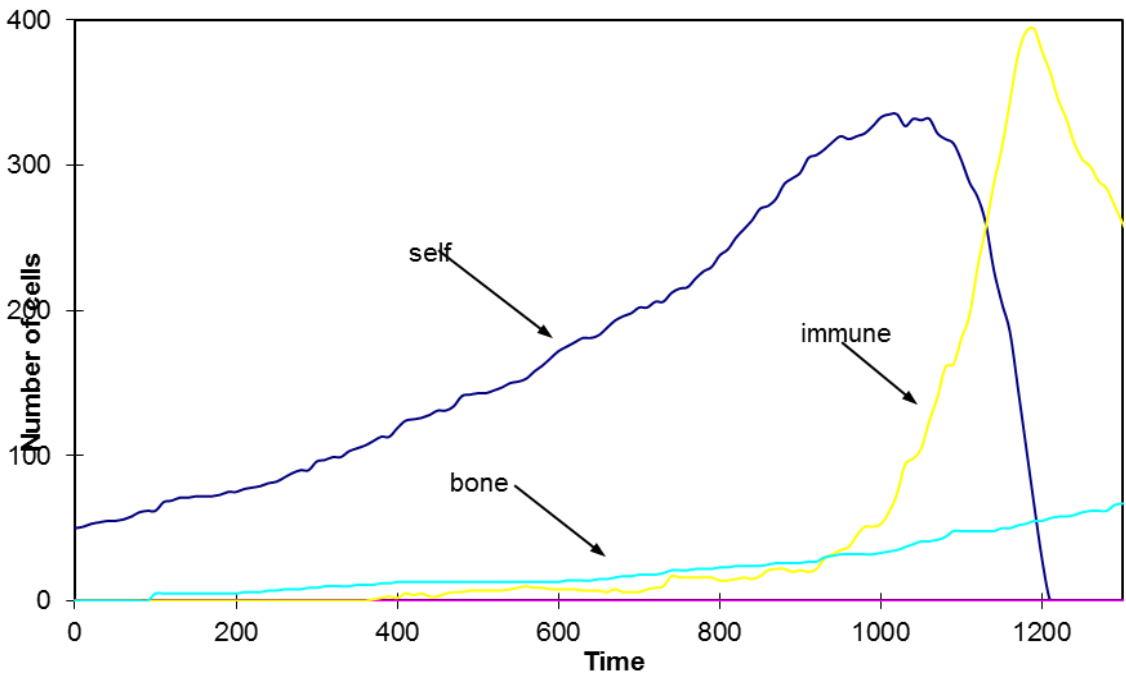

**Figure 8.**

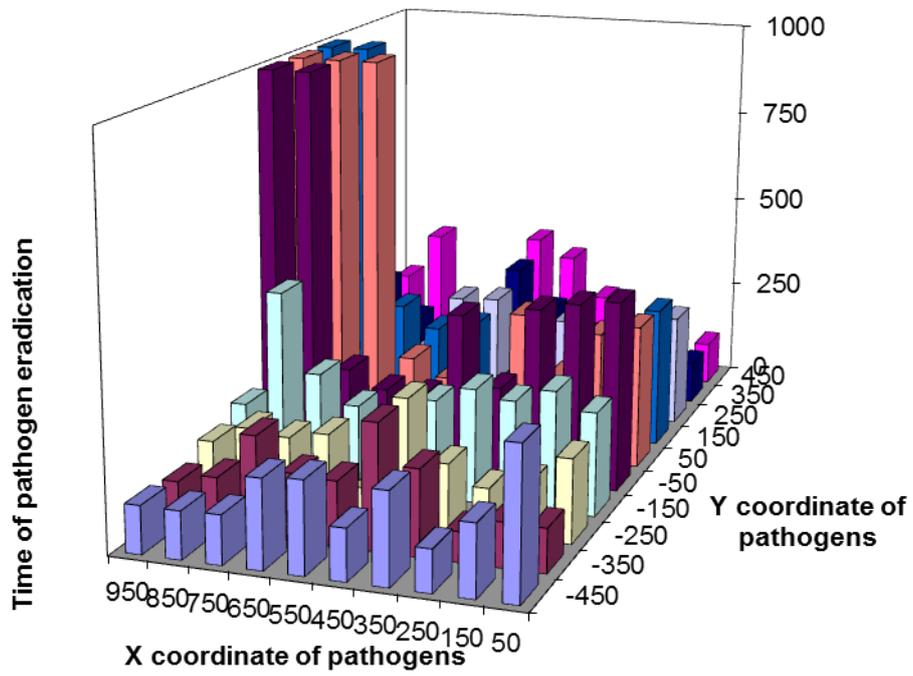

**Figure 9.**

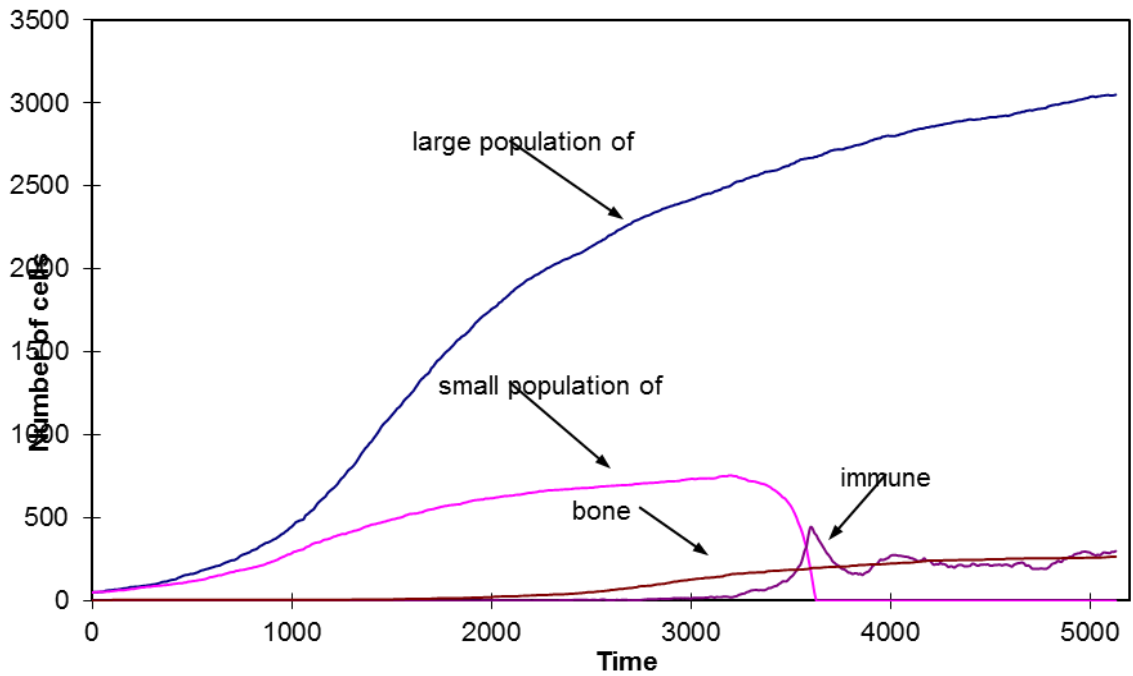

**Figure 10.**

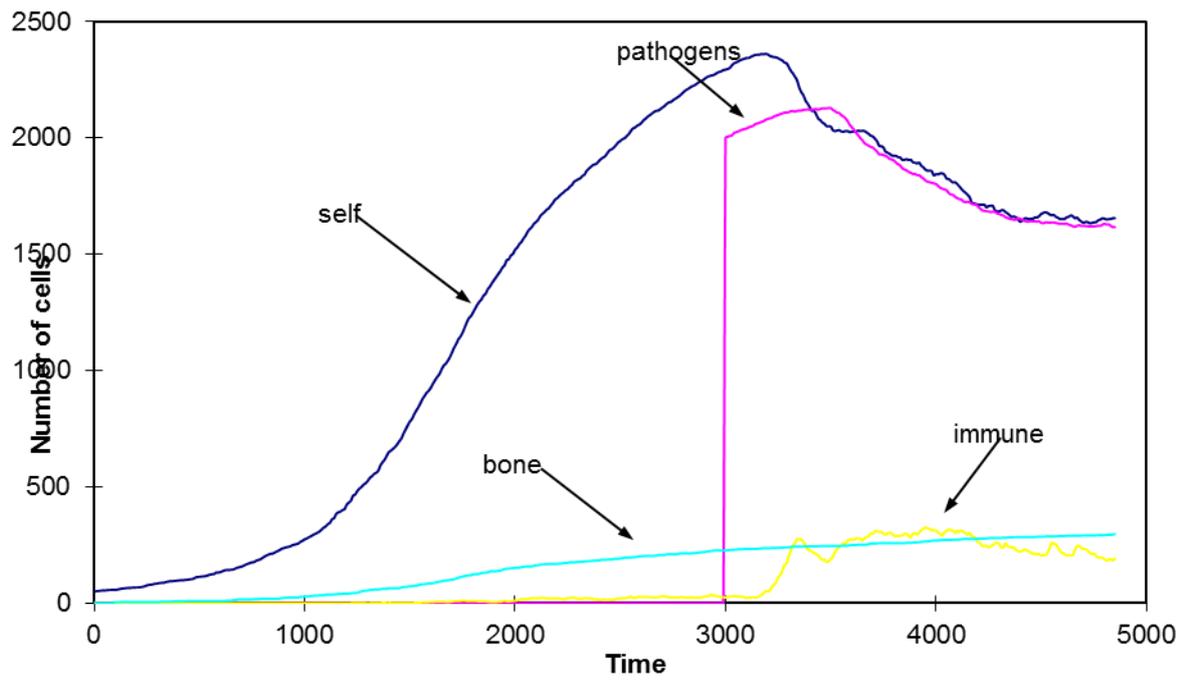